# Green-Schwarz, Nambu-Goto Actions, and Cayley's Hyperdeterminant

Hitoshi Nishino[1)] and Subhash Rajpoot[2)]

*Department of Physics & Astronomy*
*California State University*
*1250 Bellflower Boulevard*
*Long Beach, CA 90840*

## Abstract

It has been recently shown that Nambu-Goto action can be re-expressed in terms of Cayley's hyperdeterminant with the manifest $SL(2,\mathbb{R}) \times SL(2,\mathbb{R}) \times SL(2,\mathbb{R})$ symmetry. In the present paper, we show that the same feature is shared by Green-Schwarz $\sigma$-model for $N=2$ superstring whose target space-time is $D=2+2$. When its zweibein field is eliminated from the action, it contains the Nambu-Goto action which is nothing but the square root of Cayley's hyperdeterminant of the pull-back in superspace $\sqrt{\mathcal{D}\text{et}\,(\Pi_{i\alpha\dot\alpha})}$ manifestly invariant under $SL(2,\mathbb{R}) \times SL(2,\mathbb{R}) \times SL(2,\mathbb{R})$. The target space-time $D=2+2$ can accommodate self-dual supersymmetric Yang-Mills theory. Our action has also fermionic $\kappa$-symmetry, satisfying the criterion for its light-cone equivalence to Neveu-Schwarz-Ramond formulation for $N=2$ superstring.

[1)] E-Mail: hnishino@csulb.edu
[2)] E-Mail: rajpoot@csulb.edu

1. Introduction

Cayley's hyperdeterminant [1], initially an object of mathematical curiosity, has found its way in many applications to physics [2]. For instance, it has been used in the discussions of quantum information theory [3][4], and the entropy of the STU black hole [5][6] in four-dimensional string theory [7].

More recently, it has been shown [8] that Nambu-Goto (NG) action [9][10] with the $D = 2+2$ target space-time possesses the manifest global $SL(2, \mathbb{R}) \times SL(2, \mathbb{R}) \times SL(2, \mathbb{R}) \equiv [SL(2, \mathbb{R})]^3$ symmetry. In particular, the square root of the determinant of an inner product of pull-backs can be rewritten exactly as a Cayley's hyperdeterminant [1] realizing the manifest $[SL(2, \mathbb{R})]^3$ symmetry.

It is to be noted that the space-time dimensions $D = 2+2$ pointed out in [8] are nothing but the consistent target space-time of $N = 2$[3]) NSR superstring [16][17][18][19][13][14][15]. However, the NSR formulation [16][17] has a drawback for rewriting it purely in terms of a determinant, due to the presence of fermionic superpartners on the 2D world-sheet. On the other hand, it is well known that a GS formulation [12] without explicit world-sheet supersymmetry is classically equivalent to a NSR formulation [11] on the light-cone, when the former has fermionic $\kappa$-symmetry [20][15]. From this viewpoint, a GS $\sigma$-model formulation in [14] of $N = 2$ superstring [16][17][13] seems more advantageous, despite the temporary sacrifice of world-sheet supersymmetry. However, even the GS formulation [14] itself has an obstruction, because obviously the kinetic term in the GS action is not of the NG-type equivalent to a Cayley's hyperdeterminant.

In this paper, we overcome this obstruction, by eliminating the zweibein (or 2D metric) *via* its field equation which is *not* algebraic. Despite the *non*-algebraic field equation, such an elimination is possible, just as a NG action [9][10] is obtained from a Polyakov action [21]. Similar formulations are known to be possible for Type I, heterotic, or Type II superstring theories, but here we need to deal with $N = 2$ superstring [16] with the target space-time $D = 2+2$ instead of 10D. We show that the same global $[SL(2, \mathbb{R})]^3$ symmetry [8] is inherent also in $N=2$ GS action in [14] with $N=(1,1)$ supersymmetry in $D=2+2$ as the special case of [13], when the zweibein field is eliminated from the original action, re-expressed in terms of NG-type determinant form.

---

[3]) The $N = 2$ here implies the number of world-sheet supersymmetries in the Neveu-Schwarz-Ramond (NSR) formulation [11]. Its corresponding Green-Schwarz (GS) formulation [12][13][14] might be also called '$N = 2$' GS superstring in the present paper. Needless to say, the number of world-sheet supersymmetries should *not* be confused with that of space-time supersymmetries, such as $N = 1$ for Type I superstring, or $N = 2$ for Type IIA or IIB superstring [15].

As is widely recognized, the quantum-level equivalence of NG action [9][10] to Polyakov action [21] has not been well established even nowadays [22]. As such, we do not claim the quantum equivalence of our formulation to the conventional $N=2$ NSR superstring [16][17] or even to $N=2$ GS string [13] itself. In this paper, we point out only the existence of fermionic $\kappa$-symmetry and the manifest global $[SL(2, \mathbb{R})]^3$ symmetry with Cayley's hyperdeterminant as classical-level symmetries, after the elimination of 2D metric from the classical GS action [14] of $N=2$ superstring [16][17].

As in $N=2$ NSR superstring [16][17], the target $D = (2,2;2,2)^{4)}$ superspace [19] of $N=2$ GS superstring [14] can accommodate self-dual supersymmetric Yang-Mills (SDSYM) multiplet [18][19] with $N=(1,1)$ space-time supersymmetry [13][19][14], which is supersymmetric generalization of purely bosonic YM theory in $D=2+2$ [23]. The importance of the latter is due to the conjecture [24] that all the bosonic integrable or soluble models in dimensions $D \leq 3$ are generated by self-dual Yang-Mills (SDYM) theory [23]. Then it is natural to 'supersymmetrize' this conjecture [24], such that all the supersymmetric integrable models in $D \leq 3$ are generated by SDSYM in $D=2+2$ [18][19], and thereby the importance of $N=2$ GS $\sigma$-model in [14] is also re-emphasized.

In the next two sections, we present our total action of $N=2$ GS $\sigma$-model [14] whose target superspace is $D=(2,2;2,2)$ [19], and show the existence of fermionic $\kappa$-symmetry [20] as well as $[SL(2, \mathbb{R})]^3$ symmetry, due to the Cayley's hyperdeterminant for the kinetic terms in the NG form. We next confirm that our action is derivable from the $N=2$ GS $\sigma$-model [14] which is light-cone equivalent to $N=2$ NSR superstring [16][17], by elimi-

---

[4)] We use in this paper the symbol $D = (2,2;2,2)$ for the target superspace, meaning $2+2$ bosonic coordinates, plus 2 chiral and 2 anti-chiral fermionic coordinates [19][14]. In terms of supersymmetries in the *target* $D=2+2$ space-time, this superspace corresponds to $N=(1,1)$ [19][14], which should not be confused with $N=2$ on the world-sheet. In other words, $D=(2,2;2,2)$ is superspace for $N=(1,1)$ supersymmetry realized on $D=2+2$ space-time. Maximally, we can think of $N=(4,4)$ supersymmetry for SDSYM [18], but we focus only on $N=(1,1)$ supersymemtry in this paper.

nating a zweibein or a 2D metric.

## 2. Total Action with $[SL(2,\mathbb{R})]^3$ Symmetry

We first give our total action with manifest global $[SL(2,\mathbb{R})]^3$ symmetry, then show its fermionic $\kappa$-symmetry [20]. Our action has classical equivalence to the GS $\sigma$-model formulation [14] of $N=2$ superstring [16][17] with the right $D=(2,2;2,2)$ target superspace that accommodates self-dual supersymmetric YM multiplet [17][19][18][14]. In this section, we first give our total action of our formulation, leaving its derivation or justifications for later sections.

Our total action $I \equiv \int d^2\sigma \mathcal{L}$ has the fairly simple lagrangian

$$\mathcal{L} = +\sqrt{-\det(\Gamma_{ij})} + \epsilon^{ij}\Pi_i{}^A \Pi_j{}^B B_{BA} \tag{2.1a}$$

$$= +\sqrt{+\mathcal{D}\text{et}(\Pi_{i\alpha\underline{\alpha}})}\left(1 + 2\Pi_-{}^A \Pi_+{}^B B_{BA}\right) \equiv \mathcal{L}_{\text{NG}} + \mathcal{L}_{\text{WZNW}} \ , \tag{2.1b}$$

where respectively the two terms $\mathcal{L}_{\text{NG}}$ and $\mathcal{L}_{\text{WZNW}}$ are called 'NG-term' and 'WZNW-term'. The indices $i, j, \cdots = 0, 1$ are for the curved coordinates on the 2D world-sheet, while $+, -$ are for the light-cone coordinates for the local Lorentz frames, respectively defined by the projectors

$$P_{(i)}{}^{(j)} \equiv \tfrac{1}{2}(\delta_{(i)}{}^{(j)} + \epsilon_{(i)}{}^{(j)}) \ , \quad Q_{(i)}{}^{(j)} \equiv \tfrac{1}{2}(\delta_{(i)}{}^{(j)} - \epsilon_{(i)}{}^{(j)}) \ , \tag{2.2}$$

where $(i), (j), \cdots = (0), (1), \cdots$ are used for local Lorentz coordinates, and $(\eta_{(i)(j)}) = \text{diag.}(+,-)$. Note that $\delta_+{}^+ = \delta_-{}^- = +1$, $\epsilon_+{}^+ = -\epsilon_-{}^- = +1$, $\eta_{++} = \eta_{--} = 0$, $\eta_{+-} = \eta_{-+} = 1$. Whereas $\Pi_i{}^A$ is the superspace pull-back, $\Gamma_{ij}$ is a product of such pull-backs:

$$\Pi_i{}^A \equiv (\partial_i Z^M) E_M{}^A \ , \tag{2.3a}$$

$$\Gamma_{ij} \equiv \eta_{\underline{ab}} \Pi_i{}^{\underline{a}} \Pi_j{}^{\underline{b}} = \Pi_i{}^{\underline{a}} \Pi_{j\underline{a}} \ , \tag{2.3b}$$

for the target superspace coordinates $Z^M$. The $(\eta_{\underline{ab}}) = \text{diag.}(+,+,-,-)$ is the $D = 2+2$ space-time metric. We use the indices $\underline{a}, \underline{b}, \cdots = 0, 1, 2, 3$ (or $\underline{m}, \underline{n}, \cdots = 0, 1, 2, 3$) for the bosonic local Lorentz (or curved) coordinates. The $E_M{}^A$ is the flat background vielbein [25] for $D = (2,2;2,2)$ target superspace [19][14]. Its explicit form is

$$(E_M{}^A) = \begin{pmatrix} \delta_{\underline{m}}{}^{\underline{a}} & 0 \\ -\tfrac{i}{2}(\sigma^{\underline{a}}\theta)_{\underline{\mu}} & \delta_{\underline{\mu}}{}^{\underline{\alpha}} \end{pmatrix} \ , \quad (E_A{}^M) = \begin{pmatrix} \delta_{\underline{a}}{}^{\underline{m}} & 0 \\ +\tfrac{i}{2}(\sigma^{\underline{m}}\theta)_{\underline{\alpha}} & \delta_{\underline{\alpha}}{}^{\underline{\mu}} \end{pmatrix} \ . \tag{2.4}$$

We use the underlined Greek indices: $\underline{\alpha} \equiv (\alpha,\dot{\alpha})$, $\underline{\beta} \equiv (\beta,\dot{\beta})$, $\cdots$ for the pair of fermionic indices, where $\alpha, \beta, \cdots = 1, 2$ are for chiral coordinates, and $\dot{\alpha}, \dot{\beta}, \cdots = \dot{1}, \dot{2}$ are for anti-chiral coordinates [19]. The indices $\mu, \nu, \cdots = 1, 2, 3, 4$ are for curved fermionic coordinates. Similarly to the superspace for the Minkowski space-time with the signature $(+,-,-,-)$ [25], a bosonic index is equivalent to a pair of fermionic indices, e.g., $\Pi_i{}^{\underline{a}} \equiv \Pi_i{}^{\alpha\dot{\alpha}}$. In (2.4), we use the expressions like $(\sigma^{\underline{a}}\theta)_{\underline{\alpha}} \equiv -(\sigma^{\underline{a}})_{\underline{\alpha}\underline{\beta}}\theta^{\underline{\beta}}$ for the $\sigma$-matrices in $D = 2+2$ [26][19]. Relevantly, the only non-vanishing supertorsion components are [19][14]

$$T_{\underline{\alpha}\underline{\beta}}{}^{\underline{c}} = i(\sigma^{\underline{c}})_{\underline{\alpha}\underline{\beta}} = \begin{cases} +i(\sigma_{\underline{c}})_{\alpha\dot{\beta}} \ , \\ +i(\sigma_{\underline{c}})_{\dot{\alpha}\beta} = +i(\sigma_{\underline{c}})_{\beta\dot{\alpha}} \ . \end{cases} \quad (2.5)$$

The antisymmetric tensor superfield $B_{AB}$ has the superfield strength

$$G_{ABC} \equiv \tfrac{1}{2}\nabla_{[A}B_{BC)} - \tfrac{1}{2}T_{[AB|}{}^D B_{D|C)} \ . \quad (2.6)$$

Our anti-symmetrization rule is such as $M_{[AB)} \equiv M_{AB} - (-1)^{AB}M_{BA}$ *without* the factor $1/2$. The flat-background values of $G_{ABC}$ is [19][14]

$$G_{\underline{\alpha}\underline{\beta}\underline{c}} = +\tfrac{i}{2}(\sigma_{\underline{c}})_{\underline{\alpha}\underline{\beta}} = \begin{cases} +\tfrac{i}{2}(\sigma_{\underline{c}})_{\alpha\dot{\beta}} \ , \\ +\tfrac{i}{2}(\sigma_{\underline{c}})_{\dot{\alpha}\beta} = +\tfrac{i}{2}(\sigma_{\underline{c}})_{\beta\dot{\alpha}} \ . \end{cases} \quad (2.7)$$

In our formulation, the lagrangian (2.1a) needs the 'square root' of the matrix $\Gamma_{ij}$, analogous to the zweibein $e_i{}^{(j)}$ as the 'square root' of the 2D metric $g_{ij}$, defined by

$$\gamma_i{}^{(k)}\gamma_{j(k)} = \Gamma_{ij} \ , \qquad \gamma_{(k)}{}^i \gamma^{(k)j} = \Gamma^{ij} \ , \quad (2.8a)$$

$$\gamma_i{}^{(k)}\gamma_{(k)}{}^j = \delta_i{}^j \ , \qquad \gamma_{(i)}{}^k \gamma_k{}^{(j)} = \delta_{(i)}{}^{(j)} \ . \quad (2.8b)$$

Relevantly, we have $\gamma = \sqrt{-\Gamma}$ for $\Gamma \equiv \det(\Gamma_{ij})$ and $\gamma \equiv \det(\gamma_i{}^{(j)})$. We define $\Pi_\pm{}^A \equiv \gamma_\pm{}^i \Pi_i{}^A$ for the $\pm$ local light-cone coordinates. For our formulation with (2.1), we always use the $\gamma$'s to convert the curved indices $i, j, \cdots = 0, 1$ into local Lorentz indices $(i), (j), \cdots = (0), (1)$.

From (2.8), it is clear that we can always define the 'square root' of $\Gamma_{ij}$ of (2.3b) just as we can always define the zweibein $e_i{}^{(j)}$ out of a 2D metric $g_{ij}$. In fact, (2.8) determines $\gamma_i{}^{(j)}$ up to 2D local Lorentz transformations $O(1,1)$, because (2.8) is covariant under arbitrary $O(1,1)$. However, (2.8) has much more significance, because if the curved indices $ij$ of $\Gamma_{ij}$ are converted into 'local' ones, then it amounts to

$$\begin{aligned}\Gamma_{(i)(j)} &= \gamma_{(i)}{}^k\gamma_{(j)}{}^l\Gamma_{kl} = \gamma_{(i)}{}^k\gamma_{(j)}{}^l(\gamma_k{}^{(m)}\gamma_{l(m)}) \\ &= (\gamma_{(i)}{}^k\gamma_k{}^{(m)})(\gamma_{(j)}{}^l\gamma_{l(m)}) = \delta_{(i)}{}^{(m)}\eta_{(j)(m)} = \eta_{(i)(j)} \quad\Longrightarrow\quad \Gamma_{(i)(j)} = \eta_{(i)(j)} \ . \end{aligned} \quad (2.9)$$

In terms of light-cone coordinates, this implies formally the Virasoro conditions [27]

$$\Gamma_{++} \equiv \Pi_+{}^{\underline{a}}\Pi_{+\underline{a}} = 0 \ , \quad \Gamma_{--} \equiv \Pi_-{}^{\underline{a}}\Pi_{-\underline{a}} = 0 \ , \quad (2.10)$$

because $\eta_{++} = \eta_{--} = 0$. The only caveat here is that our $\gamma_i{}^{(j)}$ is not exactly the zweibein $e_i{}^{(j)}$, but it differs only by certain factor, as we will see in (4.6).

The result (2.10) is not against the original results in NG formulation [9][10]. At first glance, since the NG action has no metric, it seems that Virasoro condition [27] will not follow, unless a 2D metric is introduced as in Polyakov formulation [21]. However, it has been explicitly shown that the Virasoro conditions follow as first-order constraints, when canonical quantization is performed [10]. Naturally, this quantum-level result is already reflected at the classical level, *i.e.*, the Virasoro condition (2.10) follows, when the $ij$ indices on $\Gamma_{ij} \equiv \Pi_i{}^{\underline{a}}\Pi_{j\underline{a}}$ are converted into 'local Lorentz indices' by using the $\gamma$'s in (2.8).

Most importantly, $\mathcal{D}\mathrm{et}\,(\Pi_{i\alpha\dot\alpha})$ in (2.1b) is a Cayley's hyperdeterminant [1][8], related to the ordinary determinant in (2.1a) by

$$\mathcal{D}\mathrm{et}\,(\Pi_{i\alpha\dot\alpha}) = -\tfrac{1}{2}\epsilon^{ij}\epsilon^{kl}\epsilon^{\alpha\beta}\epsilon^{\gamma\delta}\epsilon^{\dot\alpha\dot\beta}\epsilon^{\dot\gamma\dot\delta}\Pi_{i\alpha\dot\alpha}\Pi_{j\beta\dot\beta}\Pi_{k\gamma\dot\gamma}\Pi_{l\delta\dot\delta} = -\det(\Gamma_{ij}) \ , \quad (2.11\mathrm{a})$$

$$\Gamma_{ij} \equiv \Pi_i{}^{\underline{a}}\Pi_{j\underline{a}} = \Pi_i{}^{\alpha\dot\alpha}\Pi_{j\alpha\dot\alpha} = \epsilon^{\alpha\beta}\epsilon^{\dot\gamma\dot\delta}\Pi_{i\alpha\dot\gamma}\Pi_{j\beta\dot\delta} \ . \quad (2.11\mathrm{b})$$

The global $[SL(2,\mathbb{R})]^3$ symmetry of our action $I$ is more transparent in terms of Cayley's hyperdeterminant, because of its manifest invariance under $[SL(2,\mathbb{R})]^3$. For other parts of our lagrangian, consider the infinitesimal transformation for the first factor group[5] of $SL(2,\mathbb{R}) \times SL(2,\mathbb{R}) \times SL(2,\mathbb{R})$ with the infinitesimal real constant traceless 2 by 2 matrix parameter $p$ as

$$\delta_p\Pi_i{}^A = p_i{}^j\pi_j{}^A \ , \quad \delta_p\gamma_{(i)}{}^j = -p_k{}^j\gamma_{(i)}{}^k \quad (p_i{}^i = 0) \ . \quad (2.12)$$

---

[5] In a sense, this invariance is trivial, because $SL(2,\mathbb{R}) \subset GL(2,\mathbb{R})$, where the latter is the 2D general covariance group.

The latter is implied by the definition of $\Gamma_{ij} \equiv \Pi_i{}^{\underline{a}}\Pi_{j\underline{a}}$ and $\gamma_{(i)}{}^j$ in (2.8). Eventually, we have $\delta_p \Pi_{(i)}{}^A = 0$, while $\mathcal{L}_{\text{WZNW}}$ is also invariant, thanks to $\delta_p \Pi_{(i)}{}^A = 0$. This concludes $\delta_p \mathcal{L} = 0$.

The second and third factor groups in $SL(2,\mathbb{R}) \times SL(2,\mathbb{R}) \times SL(2,\mathbb{R})$ act on the fermionic coordinates $\alpha$ and $\dot{\alpha}$ in $D = (2,2;2,2)$, which need an additional care. We first need the alternative expression of $\mathcal{L}_{\text{WZNW}}$ by the use of Vainberg construction [28][29]:

$$\mathcal{L} = +\sqrt{+\mathcal{D}\text{et}\,(\Pi_{i\alpha\dot{\alpha}})} + i\int d^3\hat{\sigma}\,\hat{\epsilon}^{\hat{i}\hat{j}\hat{k}}\,\widehat{\Pi}_{\hat{i}\alpha\dot{\alpha}}\,\widehat{\Pi}_{\hat{j}}{}^\alpha\,\widehat{\Pi}_{\hat{k}}{}^{\dot{\alpha}}\ . \tag{2.13}$$

We need this alternative expression, because superfield strength $G_{ABC}$ is less ambiguous than its potential superfield $B_{AB}$ avoiding the subtlety with the indices $\alpha$ and $\dot{\alpha}$. In the Vainberg construction [28][29], we are considering the extended 3D 'world-sheet' with the coordinates $(\hat{\sigma}^{\hat{i}}) \equiv (\sigma^i, y)$ ($\hat{i} = 0,\,1,\,2$), where $\hat{\sigma}^2 \equiv y$ is a new coordinate with the range $0 \leq y \leq 1$. Relevantly, $\hat{\epsilon}^{\hat{i}\hat{j}\hat{k}}$ is totally antisymmetric constant, and $\hat{\epsilon}^{2\hat{i}\hat{j}} = \epsilon^{ij}$. All the *hatted* indices and quantities refer to the new 3D. Any *hatted* superfield as a function of $\hat{\sigma}^i$ should satisfy the conditions [28], *e.g.*,

$$\widehat{Z}^M(\sigma, y=1) = Z^M(\sigma)\ , \qquad \widehat{Z}^M(\sigma, y=0) = 0\ . \tag{2.14}$$

Consider next the isomorphism $SL(2,\mathbb{R}) \approx Sp(1)$ [30] for the last two groups in $SL(2,\mathbb{R}) \times SL(2,\mathbb{R}) \times SL(2,\mathbb{R}) \approx SL(2,\mathbb{R}) \times Sp(1) \times Sp(1)$. These two $Sp(1)$ groups are acting respectively on the spinorial indices $\alpha$ and $\dot{\alpha}$. The contraction matrices $\epsilon_{\alpha\beta}$ and $\epsilon_{\dot{\alpha}\dot{\beta}}$ are the metrics of these two $Sp(1)$ groups, used for raising/lowering these spinorial indices. Now the infinitesimal transformation parameters of $Sp(1) \times Sp(1)$ can be 2 by 2 real constant symmetric matrices $q_{\alpha\beta}$ and $r_{\dot{\alpha}\dot{\beta}}$ acting as

$$\delta_q \widehat{\Pi}_{\hat{i}\alpha} = -q^\alpha{}_\beta \widehat{\Pi}_{\hat{i}}{}^\beta\ , \qquad \delta_q \widehat{\Pi}_{\hat{i}\alpha\dot{\alpha}} = q_\alpha{}^\gamma \widehat{\Pi}_{\hat{i}\gamma\dot{\alpha}}\ , \tag{2.15a}$$

$$\delta_r \widehat{\Pi}_{\hat{i}}{}^{\dot{\alpha}} = -r^{\dot{\alpha}}{}_{\dot{\beta}} \widehat{\Pi}_{\hat{i}}{}^{\dot{\beta}}\ , \qquad \delta_r \widehat{\Pi}_{\hat{i}\alpha\dot{\alpha}} = r_{\dot{\alpha}}{}^{\dot{\gamma}} \widehat{\Pi}_{\hat{i}\alpha\dot{\gamma}}\ , \tag{2.15b}$$

where $q^\alpha{}_\beta \equiv \epsilon^{\alpha\gamma} q_{\gamma\beta}$, $r^{\dot{\alpha}}{}_{\dot{\beta}} \equiv \epsilon^{\dot{\alpha}\dot{\gamma}} r_{\dot{\gamma}\dot{\beta}}$, *etc.* Then it is easy to confirm for $\mathcal{L}_{\text{WZNW}}$ that

$$\delta_q \left(\widehat{\Pi}_{\hat{i}\alpha\dot{\alpha}}\,\widehat{\Pi}_{\hat{j}}{}^\alpha\,\widehat{\Pi}_{\hat{k}}{}^{\dot{\alpha}}\right) = 0\ , \qquad \delta_r \left(\widehat{\Pi}_{\hat{i}\alpha\dot{\alpha}}\,\widehat{\Pi}_{\hat{j}}{}^\alpha\,\widehat{\Pi}_{\hat{k}}{}^{\dot{\alpha}}\right) = 0\ , \tag{2.16}$$

because of $q_\alpha{}^\gamma = +q^\gamma{}_\alpha$ and $r_{\dot\alpha}{}^{\dot\gamma} = +r^{\dot\gamma}{}_{\dot\alpha}$. We thus have the total invariances $\delta_q \mathcal{L} = 0$ and $\delta_r \mathcal{L} = 0$. Since $\delta_p \mathcal{L} = 0$ has been confirmed after (2.12), this concludes the $[SL(2, \mathbb{R})]^3$-invariance proof of our action (2.1).

It was pointed out in ref. [8] that 'hidden' discrete symmetry also exists in NG-action under the interchange of the three indices for $[SL(2, \mathbb{R})]^3$. In our system, however, this hidden triality seems absent. This can be seen in (2.1b), where the Cayley's hyperdeterminant or $\mathcal{L}_{\mathrm{NG}}$ indeed possesses the discrete symmetry for the three indices $i\ \alpha\ \dot\alpha$, while it is lost in $\mathcal{L}_{\mathrm{WZNW}}$. This is because the mixture of $\Pi_{i\alpha\dot\alpha}$ and $\Pi_i{}^\alpha$ or $\Pi_i{}^{\dot\alpha}$ via the non-zero components of $B_{AB}$ breaks the exchange symmetry among $i\ \alpha\ \dot\alpha$, *unlike* Cayley's hyperdeterminant.

## 3. Fermionic Invariance of our Action

We now discuss our fermionic $\kappa$-invariance. Our action (2.1) is invariant under

$$(\delta_\kappa Z^M) E_M{}^{\underline{\alpha}} = +i(\sigma_{\underline{b}})_{\underline{\alpha}}{}^{\underline{\beta}} \kappa_{-\underline{\beta}} \Pi_+{}^{\underline{b}} \equiv +i(\slashed{\Pi}_+ \kappa_-)^{\underline{\alpha}} \ , \tag{3.1a}$$

$$(\delta_\kappa Z^M) E_M{}^{\underline{a}} = 0 \ , \tag{3.1b}$$

$$\delta_\kappa \Gamma_{ij} = +\left[\kappa_-{}^{\underline{\alpha}} (\sigma_{\underline{a}} \sigma_{\underline{c}})_{\underline{\alpha}}{}^{\underline{\beta}} \Pi_{(j|\underline{\beta}}\right] \Pi_+{}^{\underline{a}} \Pi_{|i)}{}^{\underline{c}} \equiv +(\overline{\kappa}_- \slashed{\Pi}_+ \slashed{\Pi}_{(i} \Pi_{j)}) \ . \tag{3.1c}$$

The $\kappa_-{}^{\underline{\alpha}}$ is the parameter for our fermionic symmetry transformation, just as in the conventional Green-Schwarz superstring [12][20]. Since $Z^M$ is the only fundamental field in our formulation, (3.1c) is the necessary condition of (3.1a) and (3.1b).

We can confirm $\delta_\kappa I = 0$ easily, once we know the intermediate results:

$$\delta_\kappa \mathcal{L}_{\mathrm{NG}} = +\sqrt{-\Gamma} (\overline{\kappa}_- \slashed{\Pi}_+ \slashed{\Pi}_{(i} \Pi^{(i)}) \ , \tag{3.2a}$$

$$\delta_\kappa \mathcal{L}_{\mathrm{WZNW}} = -\epsilon^{ij} (\overline{\kappa}_- \slashed{\Pi}_+ \slashed{\Pi}_i \Pi_j) \ . \tag{3.2b}$$

By using the relationships, such as $\sqrt{-\Gamma} \epsilon^{(k)(l)} = +\epsilon^{ij} \gamma_i{}^{(k)} \gamma_j{}^{(l)}$, with the most crucial equation (2.10), we can easily confirm that the sum (3.2a) + (3.2b) vanishes:

$$\delta_\kappa \mathcal{L} = \delta_\kappa (\mathcal{L}_{\mathrm{NG}} + \mathcal{L}_{\mathrm{WZNW}}) = +2\sqrt{-\Gamma}\, (\overline{\kappa}_- \Pi_-)\, \Pi_+{}^{\underline{a}} \Pi_{+\underline{a}} = 0 \ . \tag{3.3}$$

Thus the fermionic $\kappa$-invariance $\delta_\kappa I = 0$ works also in our formulation, despite the absence of the 2D metric or zweibein. The existence of fermionic $\kappa$-symmetry also guarantees the light-cone equivalence of our system to the conventional $N = 2$ GS superstring [14].

## 4. Derivation of Lagrangian and Fermionic Symmetry

In this section, we start with the conventional GS $\sigma$-model action [14] for $N = 2$ superstring [16][17], and derive our lagrangian (2.1) with the fermionic transformation rule (3.1). This procedure provides an additional justification for our formulation.

The $N = 2$ GS action $I_{\text{GS}} \equiv \int d^2\sigma \mathcal{L}_{\text{GS}}$ [14] which is light-cone equivalent to $N = 2$ NSR superstring [16][17] has the lagrangian

$$\begin{aligned}
\mathcal{L}_{\text{GS}} &= +\tfrac{1}{2}\sqrt{-g}g^{ij}\Pi_i{}^{\underline{a}}\Pi_{j\underline{a}} + \epsilon^{ij}\Pi_i{}^A\Pi_j{}^B B_{BA} \\
&= +e\Pi_+{}^{\underline{a}}\Pi_{-\underline{a}} + 2e\Pi_-{}^A\Pi_+{}^B B_{BA} ,
\end{aligned} \quad (4.1)$$

where $g \equiv \det(g_{ij})$ is for the 2D metric $g_{ij}$, while $e \equiv \det(e_i{}^{(j)}) = \sqrt{-g}$ is for the zweibein $e_i{}^{(j)}$. The action $I_{\text{GS}}$ is invariant under the fermionic transformation rule [20][15][6)]

$$\delta_\lambda E^{\underline{\alpha}} = +i(\sigma_{\underline{a}})^{\underline{\alpha\beta}}\lambda^i{}_{\underline{\beta}}\Pi_i{}^{\underline{a}} = +i(\slashed{\Pi}_i\lambda^i)^{\underline{\alpha}} , \quad (4.2\text{a})$$

$$\delta_\lambda E^{\underline{a}} = 0 , \quad (4.2\text{b})$$

$$\delta_\lambda e_-{}^i = -(\lambda_-{}^{\underline{\alpha}}\Pi_{-\underline{\alpha}})e_+{}^i \equiv -(\overline{\lambda}_-\Pi_-)e_+{}^i , \quad (4.2\text{c})$$

$$\delta_\lambda e_+{}^i = 0 , \quad (4.2\text{d})$$

where $\lambda$ has only the negative component: $\lambda_{(i)}{}^{\underline{\alpha}} \equiv Q_{(i)}{}^{(j)}\lambda_{(j)}{}^{\underline{\alpha}}$. Only in this section, the local Lorentz indices are related to curved ones through the zweibein as in $\Pi_{(i)}{}^A \equiv e_{(i)}{}^j\Pi_j{}^A$, *instead of* $\gamma_i{}^{(j)}$ in the last section. In the routine confirmation of $\delta_\lambda \mathcal{L}_{\text{GS}} = 0$, we see its parallel structures to $\delta_\kappa \mathcal{L} = 0$.

We next derive our lagrangians $\mathcal{L}_{\text{NG}}$ and $\mathcal{L}_{\text{WZNW}}$ from $\mathcal{L}_{\text{GS}}$ in (4.1). To this end, we first get the 2D metric field equation from $I_{\text{GS}}{}^{7)}$

$$g_{ij} \doteq +2(g^{kl}\Pi_k{}^{\underline{b}}\Pi_{l\underline{b}})^{-1}(\Pi_i{}^{\underline{a}}\Pi_{j\underline{a}}) \equiv 2\Omega^{-1}\Gamma_{ij} \equiv h_{ij} , \quad (4.3\text{a})$$

$$\Omega \equiv g^{ij}\Pi_i{}^{\underline{a}}\Pi_{j\underline{a}} = g^{ij}\Gamma_{ij} . \quad (4.3\text{b})$$

As is well-known in string $\sigma$-models, this field equation is *not* algebraic for $g_{ij}$, because the r.h.s. of (4.3) again contains $g^{ij}$ *via* the factor $\Omega$. Nevertheless, we can formally delete the

---

[6)] We use the parameter $\lambda$ instead of $\kappa$ due to a slight difference of $\lambda$ from our $\kappa$ (Cf. eq. (4.8)).

[7)] We use the symbol $\doteq$ for a field equation to be distinguished from an algebraic one.

metric from the original lagrangian, using a procedure similar to getting NG string [9][10] from Polyakov string [21], or NG action out of Type II superstring action [12], as

$$\tfrac{1}{2}\sqrt{-g}\,g^{ij}\Gamma_{ij} = \tfrac{1}{2}\sqrt{-g}\,\Omega \doteq \tfrac{1}{2}\sqrt{-\det(h_{ij})}\,\Omega = \tfrac{1}{2}\sqrt{-\det(2\Omega^{-1}\Gamma_{ij})}\,\Omega$$
$$= \Omega^{-1}\sqrt{-\det(\Gamma_{ij})}\,\Omega = \sqrt{-\Gamma} = \mathcal{L}_{\text{NG}} \ . \tag{4.4}$$

Thus the metric disappears completely from the resulting lagrangian, leaving only $\sqrt{-\Gamma}$ which is nothing but $\mathcal{L}_{\text{NG}}$ in (2.1). As for $\mathcal{L}_{\text{WZNW}}$, since this term is metric-independent, this is exactly the same as the second term of (4.1).

We now derive our fermionic transformation rule (3.1) from (4.2). For this purpose, we establish the on-shell relationships between $e_i{}^{(j)}$ and our newly-defined $\gamma_i{}^{(j)}$. By taking the 'square root' of (4.3a), we get the $e_i{}^{(j)}$-field equation expressed in terms of the $\Pi$'s, that we call $f_i{}^{(j)}$ which coincides with $e_i{}^{(j)}$ only *on-shell*:

$$e_i{}^{(j)} \doteq f_i{}^{(j)} = f_i{}^{(j)}(\Pi_k{}^A) \ , \tag{4.5a}$$

$$f_{i(k)}f_j{}^{(k)} = h_{ij} \ , \quad f^{(k)i}f_{(k)}{}^j = h^{ij} \ , \quad f_i{}^{(k)}f_{(k)}{}^j = \delta_i{}^j \ , \quad f_{(i)}{}^k f_k{}^{(j)} = \delta_{(i)}{}^{(j)} \ . \tag{4.5b}$$

Note that the $f$'s is proportional to the $\gamma$'s by a factor of $\sqrt{\Omega/2}$, as understood by the use of (4.3), (4.5) and (2.8):

$$e_i{}^{(j)} \doteq f_i{}^{(j)} = \sqrt{\tfrac{2}{\Omega}}\gamma_i{}^{(j)} \ , \quad e_{(i)}{}^j \doteq f_{(i)}{}^j = \sqrt{\tfrac{\Omega}{2}}\gamma_{(i)}{}^j \ . \tag{4.6}$$

Recall that the factor $\Omega$ contains the 2D metric or zweibein which might be problematic in our formulation, while $\gamma_i{}^{(j)}$, $\gamma_{(i)}{}^j$ are expressed only in terms of the $\Pi_i{}^A$'s. Fortunately, we will see that $\Omega$ disappears in the end result.

Our fermionic transformation rule (3.1a) is now obtained from (4.2a), as

$$\delta_\lambda E^{\underline{\alpha}} = i(\slashed{\Pi}_i\lambda^i)^{\underline{\alpha}} \doteq if^{(i)j}(\slashed{\Pi}_j\lambda_{(i)})^{\underline{\alpha}} = i\sqrt{\tfrac{\Omega}{2}}\gamma^{(i)j}\left(\slashed{\Pi}_j\lambda_{(i)}\right)^{\underline{\alpha}}$$
$$= i\gamma^{(i)j}\left[\slashed{\Pi}_j\left(\sqrt{\tfrac{\Omega}{2}}\lambda_{(i)}\right)\right]^{\underline{\alpha}} = i(\slashed{\Pi}^{(i)}\kappa_{(i)})^{\underline{\alpha}} = \delta_\kappa E^{\underline{\alpha}} \ , \tag{4.7}$$

where $\lambda$ and $\kappa$ are proportional to each other by

$$\kappa_{(i)} \equiv \sqrt{\tfrac{\Omega}{2}}\lambda_{(i)} \ . \tag{4.8}$$

Such a re-scaling is always possible, due to the arbitrariness of the parameter $\lambda$ or $\kappa$.

As an additional consistency confirmation, we can show the $\kappa$-invariance of (2.10), using the convenient lemmas

$$(\delta_\kappa \gamma_+{}^i)\gamma_i{}^+ = (\delta_\kappa \gamma_-{}^i)\gamma_i{}^- = \tfrac{1}{2}\Omega^{-1}\delta_\kappa\Omega \ , \quad (\delta_\kappa \gamma_+{}^i)\gamma_i{}^- = 0 \ , \quad (\delta_\kappa \gamma_-{}^i)\gamma_i{}^+ = -(\overline{\kappa}_- \Pi_-) \ . \quad (4.9)$$

Combining these with (3.1c), we can easily confirm that $\delta_\kappa \Gamma_{++} = 0$ and $\delta_\kappa \Gamma_{--} = 0$, as desired for consistency of the 'built-in' Virasoro condition (2.10).

The complete disappearance of $\Omega$ in our transformation rule (3.1) is desirable, because $\Omega$ itself contains the metric that is *not* given in a closed algebraic form in terms of $\Pi_i{}^A$. If there were $\Omega$ involved in our transformation rule (3.1), it would pose a problem due to the metric $g_{ij}$ in $\Omega$. To put it differently, our action (2.1) and its fermionic symmetry (3.1) are expressed only in terms of the fundamental superfield $Z^M$ *via* $\Pi_i{}^A$ with no involvement of $g_{ij}$, $e_i{}^{(j)}$ or $\Omega$, thus indicating the total consistency of our system. This concludes the justification of our fermionic $\kappa$-transformation rule (3.1), based on the $N = 2$ GS $\sigma$-model [14] light-cone equivalent to $N = 2$ NSR superstring [16][17].

## 5. Concluding Remarks

In this paper, we have shown that after the elimination of the 2D metric at the classical level, the NG-action part $I_{\text{NG}}$ of GS $\sigma$-model action [14] for $N = 2$ superstring [16][17] is entirely expressed as the square root of a Cayley's hyperdeterminant with the manifest $[SL(2,\mathbb{R})]^3$ symmetry. In particular, this is valid in the presence of target superspace background in $D = (2,2;2,2)$ [19]. From this viewpoint, $N = 2$ GS $\sigma$-model [14] seems more suitable for discussing the $[SL(2,\mathbb{R})]^3$ symmetry *via* a Cayley's hyperdeterminant. We have seen that the $[SL(2,\mathbb{R})]^3$ symmetry acts on the three indices $i$, $\alpha$, $\dot{\alpha}$ carried by the pull-back $\Pi_{i\alpha\dot{\alpha}}$ in $\mathcal{D}\text{et}\,(\Pi_{i\alpha\dot{\alpha}})$ in $D = (2,2;2,2)$ superspace [19][14]. The hidden discrete symmetry pointed out in [8], however, seems absent in $N = 2$ string [17][19][14] due to the WZNW-term $\mathcal{L}_{\text{WZNW}}$.

We have also shown that our action (2.1) has the classical invariance under our fermionic $\kappa$-symmetry (3.1), despite the elimination of zweibein or 2D metric. Compared with the original $I_{\text{GS}}$ [14], our action has even simpler structure, because of the absence of the 2D metric or zweibein. Due to its fermionic $\kappa$-symmetry, we can also regard that our system is classically equivalent to NSR $N = 2$ superstring [16][17], or $N = 2$ GS superstring [13]. As an important by-product, we have confirmed that the Virasoro condition (2.10) are inherent even in the NG reformulation of $N = 2$ GS string [14] at the classical level. This is also consistent with the original result that Virasoro condition is inherent in NG string [9][10].

One of the important aspects is that our action (2.1) and the fermionic transformation rule (3.1) involve neither the 2D metric $g_{ij}$, the zweibein $e_i{}^{(j)}$, nor the factor $\Omega$ containing these fields. This indicates the total consistency of our formulation, purely in terms of superspace coordinates $Z^M$ as the fundamental independent field variables.

In this paper, we have seen that neither the 2D metric $g_{ij}$ nor the zweibein $e_i{}^{(j)}$, but the superspace pull-back $\Pi_{i\alpha\dot\alpha}$ is playing a key role for the manifest symmetry $[SL(2,\mathbb{R})]^3$ acting on the three indices $i\alpha\dot\alpha$. In particular, the combination $\Gamma_{ij} \equiv \Pi_i{}^{\underline{a}}\Pi_{j\underline{a}}$ plays a role of 'effective metric' on the 2D world-sheet. This suggests that our field variables $Z^M$ alone are more suitable for discussing the global $[SL(2,\mathbb{R})]^3$ symmetry of $N = 2$ superstring [16][17][14].

As a matter of fact, in $D = 2 + 2$ *unlike* $D = 3 + 1$, the components $\alpha$ and $\dot\alpha$ are *not* related to each other by complex conjugations [26][18][19]. Additional evidence is that the signature $D = 2 + 2$ seems crucial, because $SO(2,2) \approx SL(2,\mathbb{R}) \times SL(2,\mathbb{R})$ [30], while $SO(3,1) \approx SL(2,\mathbb{C})$ for $D = 3 + 1$ is not suitable for $SL(2,\mathbb{R})$. Thus it is more natural that the NG reformulation of $N = 2$ GS superstring [14] with the target superspace $D = (2,2;2,2)$ is more suitable for the global $[SL(2,\mathbb{R})]^3$ symmetry acting on the three independent indices $i, \alpha$ and $\dot\alpha$.

It seems to be a common feature in supersymmetric theories that certain non-manifest symmetry becomes more manifest only after certain fields are eliminated from an original lagrangian. For example, in $N = 1$ local supersymmetry in 4D, it is well-known that the $\sigma$-model Kähler structure shows up, only after all the auxiliary fields in chiral multiplets are eliminated [31]. This viewpoint justifies to use a NG-formulation with the 2D metric eliminated, instead of the original $N = 2$ GS formulation [13][14], in order to elucidate the global $[SL(2, \mathbb{R})]^3$ symmetry of the latter, *via* a Cayley's hyperdeterminant.

It has been well known that the superspace $D = (2, 2; 2, 2)$ is the natural background for SDYM multiplet [17][18][19][14]. Moreover, SDSYM theory [18][19][14] is the possible underlying theory for all the (supersymmetric) integrable systems in space-time dimensions lower than four [24]. All of these features strongly indicate the significant relationships among Cayley's hyperdeterminant [1][8], $N = 2$ superstring [16][17], or $N = 2$ GS superstring [13][14] with $D = (2, 2; 2, 2)$ target superspace [19][14], its NG reformulation as in this paper, the STU black holes [5][6], SDSYM theory in $D = 2 + 2$ [18][19][14], and supersymmetric integrable or soluble models [24][17][19][14] in dimensions $D \leq 3$.

We are grateful to W. Siegel and the referee for noticing mistakes in an earlier version of this paper.

## References

[1] A. Cayley, *'On the Theory of Linear Transformations'*, Camb. Math. Jour. **4** (1845) 193.

[2] M. Duff, *'String Triality, Black Hole Entropy and Cayley's Hyperdeterminant'*, `hep-th/0601134`.

[3] V. Coffman, J. Kundu and W. Wooters, `quant-ph/9907047`, Phys. Rev. **A61** (2000) 52306.

[4] A. Miyake and M. Wadati, *'Multiparticle Entanglement and Hyperdeterminants'*, ERATO Workshop on Quantum Information Science 2002 (September, '02, Tokyo, Japan), `quant-ph/0212146`.

[5] K. Behrndt, R. Kallosh, J. Rahmfeld, M. Shmakova and W.K. Wong, `hep-th/9608059`, Phys. Rev. **D54** (1996) 6293.

[6] R. Kallosh and A. Linde, `hep-th/0602061`, Phys. Rev. **D73** (2006) 104033.

[7] M.J. Duff, J.T. Liu and J. Rahmfeld, `hep-th/9508094`, Nucl. Phys. **B459** (1986) 125.

[8] M. Duff, `hep-th/0602160`, Phys. Lett. **641B** (2006) 335.

[9] Y. Nambu, *'Duality and Hydrodynamics'*, Lectures at the Copenhagen conference, 1970.

[10] T. Goto, Prog. Theor. Phys. **46** (1971) 1560.

[11] P. Ramond, Phys. Rev. **D3** (1971) 2415; A. Neveu and J.H. Schwarz, Nucl. Phys. **B31** (1971) 86.

[12] M. Green and J.H. Schwarz, Phys. Lett. **136B** (1984) 367.

[13] W. Siegel, `hep-th/9210008`, Phys. Rev. **D47** (1993) 2512.

[14] H. Nishino, `hep-th/9211042`, Int. Jour. Mod. Phys. **A9** (1994) 3077.

[15] M. Green, J.H. Schwarz and E. Witten, *'Superstring Theory'*, Vol. 1 & 2, Cambridge University Press 1986.

[16] M. Ademollo, L. Brink, A. D'Adda, R. D'Auria, E. Napolitano, S. Sciuto, E. Del Giudice, P. Di Vecchia, S. Ferrara, F. Gliozzi, R. Musto, R. Pettorino and J.H. Schwarz, Nucl. Phys. **B111** (1976) 77; L. Brink and J.H. Schwarz, Nucl. Phys. **B121** (1977) 285; A. Sen, Nucl. Phys. **B228** (1986) 287.

[17] H. Ooguri and C. Vafa, Mod. Phys. Lett. **A5** (1990) 1389; Nucl. Phys. **B361** (1991) 469; *ibid.* **B367** (1991) 83; H. Nishino and S.J. Gates, Jr., Mod. Phys. Lett. **A7** (1992) 2543.

[18] W. Siegel, `hep-th/9205075`, Phys. Rev. **D46** (1992) R3235; `hep-th/9207043`, Phys. Rev. **D47** (1993) 2504; `hep-th/9204005`, Phys. Rev. Lett. **69** (1992) 1493; A. Parkes, `hep-th/9203074`, Phys. Lett. **286B** (1992) 265.

[19] H. Nishino, S.J. Gates, Jr., and S.V. Ketov, `hep-th/9203080`, Phys. Lett. **307B** (1993) 331; `hep-th/9203081`, Phys. Lett. **307B** (1993) 323; `hep-th/9203078`, Phys. Lett. **297B** (1992) 99; `hep-th/9207042`, Nucl. Phys. **B393** (1993) 149.

[20] L. Brink and J.H. Schwarz, Phys. Lett. **100B** (1981) 310; W, Siegel, Phys. Lett. **128B** (1983) 397; Class. & Quant. Gr. **2** (1985) L95.

[21] A.M. Polyakov, Phys. Lett. **103B** (1981) 207 and 211.

[22] *For recent quantizations of NG string, see, e.g.,* K. Pohlmeyer, `hep-th/0206061`, Jour. Mod. Phys. **A19** (2004) 115; D. Bahns, `hep-th/0403108`, Jour. Math. Phys. **45** (2004) 4640; T. Thiemann, `hep-th/0401172`, Class. & Quant. Gr. **23** (2006) 1923.

[23] A.A. Belavin, A.M. Polyakov, A.S. Schwartz and Y.S. Tyupkin, Phys. Lett. **59B** (1975) 85; R.S. Ward, Phys. Lett. **61B** (1977) 81; M.F. Atiyah and R.S. Ward, Comm. Math. Phys. **55** (1977) 117; E.F. Corrigan, D.B. Fairlie, R.C. Yates and P. Goddard, Comm. Math. Phys. **58** (1978) 223; E. Witten, Phys. Rev. Lett. **38** (1977) 121.

[24] M.F. Atiyah, *unpublished*; R.S. Ward, Phil. Trans. Roy. Lond. **A315** (1985) 451; N.J. Hitchin, Proc. Lond. Math. Soc. **55** (1987) 59.

[25] J. Wess and J. Bagger, *'Superspace and Supergravity'*, Princeton University Press, 1992.

[26] T. Kugo and P.K. Townsend, Nucl. Phys. **B211** (1983) 157.

[27] M.A. Virasoro, Phys. Rev. **D1** (1970) 2933.

[28] M.M. Vainberg, *'Variational Methods for the Study of Non-Linear Operators'*, Holden Day, San Francisco, 1964.

[29] S.J. Gates, Jr. and H. Nishino, Phys. Lett. **173B** (1986) 46.

[30] R. Gilmore, *'Lie Groups, Lie Algebras and Some of Their Applications'*, Wiley-Interscience, 1973.

[31] E. Cremmer, B. Julia, J. Scherk, S. Ferrara, L. Girardello and P. van Nieuwenhuizen, Phys. Lett. **79B** (1978) 231; Nucl. Phys. **B147** (1979) 105; E. Cremmer, S. Ferrara, L. Girardello and A. van Proyen, Nucl. Phys. **B212** (1983) 413.